\documentclass{article}  
\usepackage[subpreambles=true]{standalone}
\usepackage{import}
\usepackage{amsmath}
\usepackage{amssymb}
\usepackage{hyperref}
\usepackage{graphicx}
\usepackage{booktabs}
\usepackage{float}
\usepackage{geometry}
\usepackage{listings}
\usepackage{enumitem}
\usepackage{cite}
\usepackage{subcaption}
\usepackage{color}
\usepackage{xr}
\usepackage{siunitx} 
\usepackage[font=footnotesize, labelfont=bf, labelformat=empty]{caption}
\makeatletter
\newcommand*{\addFileDependency}[1]{
  \typeout{(#1)}
  \@addtofilelist{#1}
  \IfFileExists{#1}{}{\typeout{No file #1.}}
}
\makeatother

\newcommand*{\myexternaldocument}[1]{%
    \externaldocument{#1}%
    \addFileDependency{#1.tex}%
    \addFileDependency{#1.aux}%
}

\myexternaldocument{appendix}

\title{AOSLO-net: A deep learning-based method for automatic segmentation of retinal microaneurysms from adaptive optics scanning laser ophthalmoscope images}

\author{Qian Zhang$^{1,\dag}$, Konstantina Sampani$^{2,3,\dag}$, Mengjia Xu$^{1,\dag}$, Shengze Cai${}^{1}$, \\
Yixiang Deng${}^{5}$, He Li${}^{5}$, Jennifer K. Sun$^{2,4,\ast}$, George Em Karniadakis${}^{1,5,\ast}$
\\
\\
\normalsize{${}^{1}$Division of Applied Mathematics, Brown University, Providence, RI 02912, USA}\\
\normalsize{${}^{2}$Beetham Eye Institute, Joslin Diabetes Center, Boston, MA 02215, USA}\\
\normalsize{${}^{3}$Department of Medicine, Harvard Medical School, Boston, MA 02115, USA}\\
\normalsize{${}^{4}$Department of Ophthalmology, Harvard Medical School, Boston, MA 02115, USA}\\
\normalsize{${}^{5}$School of Engineering, Brown University, Providence, RI 02912, USA}\\
\\
\normalsize{$^\dag$ These authors  contributed equally 
 to this work.}\\
\normalsize{$^\ast$ To whom correspondence should be addressed. } \\ \normalsize{E-mail: george\_karniadakis@brown.edu, Jennifer.Sun@joslin.harvard.edu.}

}

\date{}

\begin{document}
\maketitle
\begin{abstract}
      Microaneurysms (MAs) are one of the earliest signs of diabetic retinopathy (DR), a frequent complication of diabetes that can lead to visual impairment and blindness. Adaptive optics scanning laser ophthalmoscopy (AOSLO) provides real-time retinal images with resolution down to $\sim$2 $\mu m$ and thus allows detection of the morphologies of individual MAs, a potential marker that might dictate MA pathology and affect the progression of DR. In contrast to the numerous automatic models developed for assessing the number of MAs on fundus photographs, currently there is no high throughput image protocol available for automatic analysis of AOSLO photographs. To address this urgency, we introduce AOSLO-net, a deep neural network framework with customized training policies to automatically segment MAs from AOSLO images. We evaluate the performance of AOSLO-net using 87 DR AOSLO images and our results demonstrate that the proposed model outperforms the state-of-the-art segmentation model both in accuracy and cost and enables correct MA morphological classification.
\end{abstract}

\section*{Introduction}
Diabetic retinopathy (DR), a frequent complication of diabetes, remains the leading cause of new cases of blindness, among working-age adults in the United States\cite{CDC}, and it is expected to affect approximately 14.6 million people in United States by the year 2050\cite{NIHwebpage}. DR results in pathological alterations in both neural and microvascular structures and can damage any part of the central or peripheral retina~\cite{antonetti2021current}. Based on the Early Treatment Diabetic Retinopathy Study (ETDRS) severity scale, eyes with DR progress through a severity spectrum of DR from non-proliferative DR (NPDR)\cite{early2020grading}. Classification of these severity stages is performed by evaluating  certain clinical features of the retina, such as microaneurysms (MAs), intraretinal hemorrhages, hard exudates, cotton wool spots, intraretinal microvascular abnormalities and venous beading\cite{wilkinson2003proposed}. Advanced and vision-threatening stages of DR are characterized by neovascularization and by the presence of vessels that may leak fluid and eventually cause diabetic macular edema (DME), which is manifested as thickening of the central retinal\cite{Wong2016DiabeticR}.

Fundus photography, a conventional and non-invasive imaging modality, has been widely used since the 1960's to access the severity of DR on the basis of the presence and severity of retinal vascular lesions\cite{paing2016detection,yan2020deep,mitani2020detection}. In fundus images (see Fig.~\ref{Figure:Data_Type_Input}(a)), MAs appear as red circular dots with sharp margins whereas intraretinal hemorrhages, which also display as red dots or blots, may be larger and more irregularly shaped than MAs\cite{kwan2019imaging}. MAs and intraretinal hemorrhages are frequently not able to be distinguished on fundus images as the standard fundus photographs do not provide microscopic details of these lesions or allow determination of vascular perfusion. Indeed, the ETDRS severity scale lumps the grading of these lesions into a single ``H/MA" category. In contrast, more advanced imaging modalities, such as optical coherence tomography angiography (OCTA)~\cite{borrelli2019vivo,kaizu2020microaneurysm} and adaptive optics scanning laser ophthalmoscopy (AOSLO)~\cite{dubow2014classification,bernabeu2018estimation}, is capable of providing higher resolution details of these abnormalities and address the presence or absence of blood flow, thereby allowing differentiation hemorrhages from MAs~\cite{fenner2018advances}. AOSLO provides retinal imaging with the highest resolution of all the available retinal imaging techniques on human retina (down to the cellular level (2 $\mu m$)) and thus has been used to identify and quantify the morphology of individual MA~\cite{dubow2014classification} (see Figs.~\ref{Figure:Data_Type_Input}(b-e)) and to measure blood flow at the capillary level~\cite{de2016rapid}.

Recent OCTA and AOSLO-based studies have suggested a possible correlation between the morphology of retinal MAs and their tendency to leak fluid, rupture or form thrombus~\cite{schreur2019morphological, bernabeu2018estimation, li2020predictive}.  These findings imply that accurate identification of the shapes of MAs might be useful in the future to improve prediction of DR worsening or improvement. However, existing models for MA segmentation and classification have been trained on standard fundus photographs and therefore can only predict the number of MAs and their locations~\cite{ehlers2017automated, sreng2017automated, tavakoli2020automated, kou2019microaneurysms, murugan2019automatic, xie2020sesv}, because the resolution of standard fundus photography is not sufficient to analyse the shape of individual MAs. In contrast, the AOSLO imaging technique provides ultra-high resolution retinal images that can be used to classify MA morphologies~\cite{dubow2014classification}. To date, AOSLO retinal images have been analyzed manually by specially trained personnel~\cite{dubow2014classification,lammer2018association,bernabeu2018estimation,schreur2019morphological} as  no model has been developed to automatically process these images. 

Emerging interest in automated analysis of retinal images has been sparked by the rapidly increasing prevalence of diabetes worldwide and the consequent need for scalable approaches to screen and triage patients at risk for vision loss from DR~\cite{vujosevic2020screening,rajalakshmi2020impact,he2020artificial}.  With the recent advances in the computational power of graphics processing units (GPUs),  deep convolutional neural networks (DCNNs) have  become a widely used tool for efficient DR screening~\cite{grzybowski2019artificial}.  DCNNs are more generic compared to conventional methods that rely on hand-crafted features because the deep layers in the network act as a set of increased levels of feature extractors that can learn directly from the input images~\cite{pratt2016convolutional,fischbacher2021modular,liskowski2016segmenting}. Moreover, DCNN models are highly discriminative in automated DR severity grading and thus have achieved higher screening accuracy than conventional methods~\cite{vujosevic2020screening}. In particular, UNet, combining an encoder and a decoder to form a “U-shape” structure,  is a specially designed CNN architecture for biomedical image segmentation tasks.  UNet is very effective in few-shot prediction with only a few labeled images when combined with data augmentation~\cite{ronneberger2015u} and it has outperformed the plain DCNN in segmenting biomedical images,  particularly for those with complex subtleties~\cite{ronneberger2015u,xiao2018weighted,bao2021segmentation}. Recent development of UNet has given rise to a number of variants,  such as deformable UNet~\cite{jin2019dunet}, residual UNet~\cite{alom2019recurrent}, recurrent residual UNet~\cite{alom2019recurrent} and iterNet~\cite{li2020iternet}, which further improved the segmentation accuracy on fundus images.

We have noted several features from the AOSLO image dataset that pose challenges for the existing automatic segmentation models. As shown in Fig.~\ref{Figure:Data_Type_Input}(b), (i) the contrast between the MA body and background is low whereas the level of the background noise is high; (ii) the boundary of the MA is not clearly defined; (iii) a typical AOSLO image may contain one or multiple MAs with different shapes and sizes; (iv) there are numerous background blood vessels in the images, some of which are even at similar size as the MAs. These background vessels may interfere with the segmentation of the feeding or draining  vessels of the MAs, which are crucial to determine MA morphology. In this work, we design the first deep neural network model, or AOSLO-net (see Fig.~\ref{Figure:Flowchart}), to perform automatic segmentation of MAs from AOSLO images and quantify their shape metrics that can be used for classification of MAs into different types, such as focal bulging, saccular, fusiform, mixed saccular/fusiform, pedunculated and irregular-shaped MAs~\cite{dubow2014classification}. This model is trained and tested by using 87 AOSLO MA images with masks generated manually by trained graders, the largest published AOSLO image dataset for this kind of effort thus far. We evaluated the performance of this model by comparing the model predictions with nnUNet~\cite{isensee2021nnu}, a state-of-the-art (SOTA) UNet model whose superiority has been demonstrated for dozens of publicly available databases.

\section*{Data}
\subsection*{Demographic Information}
In this study, 87 MAs were imaged from 28 eyes of 20 subjects with varying severity of DR (56\% NPDR and 44\% PDR). Sixteen (80\%) subjects had type 1 diabetes, 7 (35\%) of the subjects are female and average age is $41.9\pm10.4$ years old, with mean diabetes duration 23.9±8.4 years, and mean HbA1c is $8.2\pm1.1$\%. Informed written consent was obtained from each participant prior to the performance of any study procedures at a single visit in Beetham Eye Institute, a tertiary referral center for diabetes care. This study adhered to the tenets of the Declaration of Helsinki and was approved by the institutional review board of the Joslin Diabetes Center.

\begin{figure}[H]
    \centering
    \includegraphics[width=1.0\linewidth]{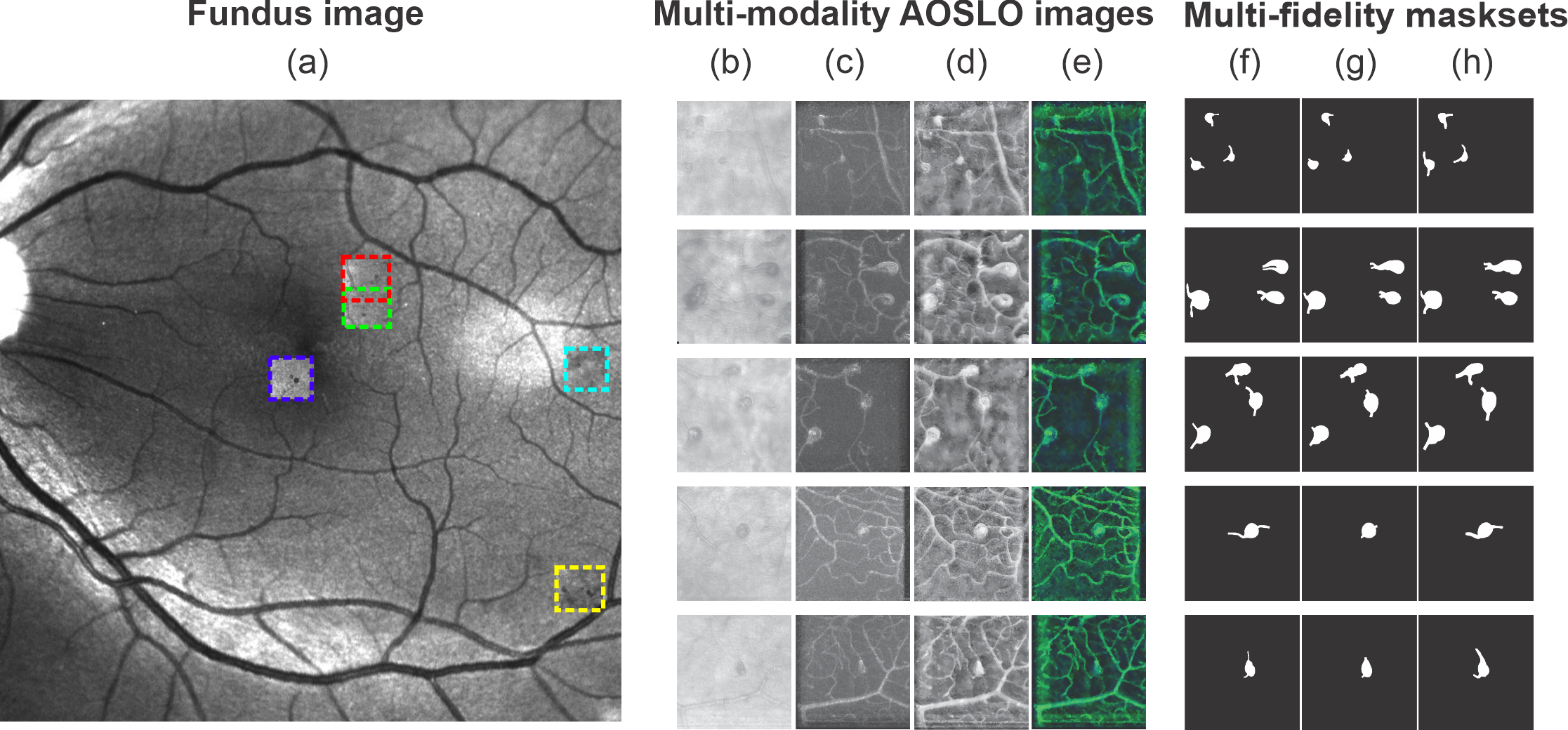}
    \caption{\textbf{Figure~\ref{Figure:Data_Type_Input}. Multi-modality AOSLO images and multi-fidelity mask sets are employed to train and test the AOSLO-net.}  (a) Five MAs imaged by AOSLO (highlighted by boxes in different colors) superimposed on a digital fundus photograph from an eye with diabetic retinopathy. (b-e) Examples of four sets of AOSLO images with different modalities employed to train and test the AOSLO-net:  (b) raw images extracted from the AOSLO videos; (c) perfusion maps (also see Fig. 1 in supporting material (SI)); (d) preprocessed AOSLO images with enhancement;  (e) two-modality images which are generated by concatenating perfusion maps (c) and enhanced AOSLO images (d). Details of how multi-modality AOSLO images are generated can be found in the Method section. These images illustrate that a varied number of MAs can be detected in a single AOSLO image. Images in row 1-3 contain multiple MAs, whereas images in row 4-5 contain a single MA with complicated background vessels whose size may be comparable to the MA. (f-h) Three sets of masks are generated independently to examine the robustness of the AOSLO-net to mask sets with different qualities. \textbf{Normal set} (f): masks are created based on both the AOSLO videos and perfusion maps to illustrate the body of MAs and their feeding and draining vessels. \textbf{Short set} (g): masks are designed to show shorter feeding and draining vessels of MAs compared to the normal set, while the thickness of the vessels remains similar to the normal masks. \textbf{Thick set} (h): masks are designed to show thicker feeding and draining vessels of MAs compared to the normal set, while the length of the vessels remains similar to the normal masks.  }\label{Figure:Data_Type_Input}
\end{figure}

\subsection*{Data Set}
87 MAs from the eyes of adult study participants with diabetes underwent AOSLO imaging. All MAs were located within $\sim$\ang{20} of the foveal center. The AOSLO system has been previously described in detail by Lu et al.~\cite{lu2016computational}. This system uses confocal and multiply scattered light (MSL) imaging modes, and achieves a field size of $\sim$\ang{1.75} $\times$~$\sim$\ang{1.75} with lateral resolution of $\sim$2.5 $\mu$m on the retina. Moreover, 75-frame videos of each MA were aligned and averaged (MATLAB, The MathWorks, Inc., Natick, MA, USA). The magnification factor on AOSLO images was determined by eye axial length measurement or derived from the spherical equivalent of the eye. For this exploratory study, we included MAs with high quality AOSLO images (see Fig.~\ref{Figure:Data_Type_Input}(b)), where the 2D MSL images and corresponding perfusion maps (see Fig.~\ref{Figure:Data_Type_Input}(c)) provided sufficient detail to identify the full extent of MAs’ bodies and their parent vessels’ boundaries. 

The masks in our dataset, as illustrated in Fig.~\ref{Figure:Data_Type_Input}(f-h), are considered as the ground truth for training the AOSLO-net and they are created manually by ophthalmologists and skilled trainees using ImageJ~\cite{schneider2012nih}. Both the AOSLO images (Fig.~\ref{Figure:Data_Type_Input}(b)) and the corresponding perfusion map (Fig.~\ref{Figure:Data_Type_Input}(c)) are referred when masks are generated. Moreover, different groups of MA masks, as shown in Figs.~\ref{Figure:Data_Type_Input}(g-h), are created by varying the length and thickness of the parent vessels to test the robustness of AOSLO-net to masks with different qualities. The normal mask set (see Fig.~\ref{Figure:Data_Type_Input}(f)), which is generated by ophthalmologists, intends to represent the true geometries of the parenting vessels illustrated on the AOSLO images. The short mask set (see Fig.~\ref{Figure:Data_Type_Input}(g)) is designed to show shorter feeding and draining vessels of MAs compared to the normal mask set, while the thickness of the vessels remains similar to the normal masks. The thick mask set (see Fig.~\ref{Figure:Data_Type_Input}(h)) is designed to show thicker feeding and draining vessels of MAs compared to the normal set, while the length of the vessels remains similar to the normal masks. This examination also demonstrates the human ability of segmentation when comparing one kind of masks with another, as they equivalently represent the MAs.

\begin{figure}[H]
    \centering
    \includegraphics[width=1.0\linewidth]{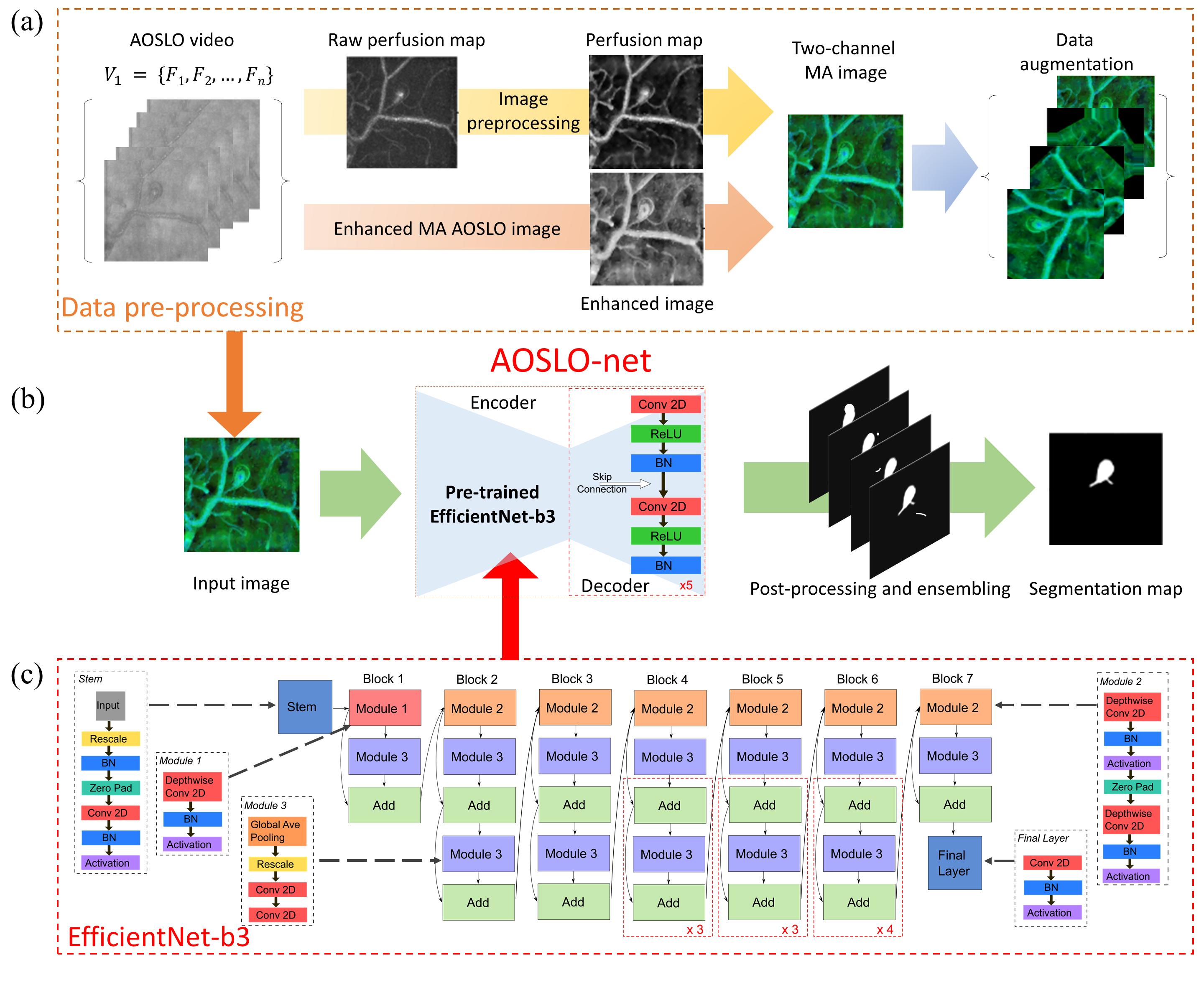}
    \caption{\textbf{Figure~\ref{Figure:Flowchart}. Overview of the architecture of AOSLO-net.} The MA segmentation is performed through the following steps: data pre-processing, deep neural network training and inference, post-processing and output ensembling. (a) Pre-processing AOSLO images for training the AOSLO-net. We first created the perfusion map by computing the deviations from the AOSLO video. Moreover, we created another set of enhanced AOSLO images from the AOSLO video. These two sets of images are concatenated to generate a third set of two-channel MA images. Data augmentation was performed on the two-channel MA images to increase the diversity of the training set (see examples in Fig. 2 in SI). (b) Detailed modular components of the AOSLO-net, i.e., preprocessed multi-modality images produced from (a) is fed into the AOSLO-net, which consists of an EfficientNet-b3 encoder (c) and the regular UNet decoder. We then perform post-processing and ensembling on the output images of the AOSLO-net to generate the segmentation map. (c) A zoom-in view of the EfficientNet-b3, which works as the encoder in the AOSLO-net.
    }\label{Figure:Flowchart}
\end{figure}

\section*{Results}
\begin{figure}[H]
    \centering
    \includegraphics[width=0.88\linewidth]{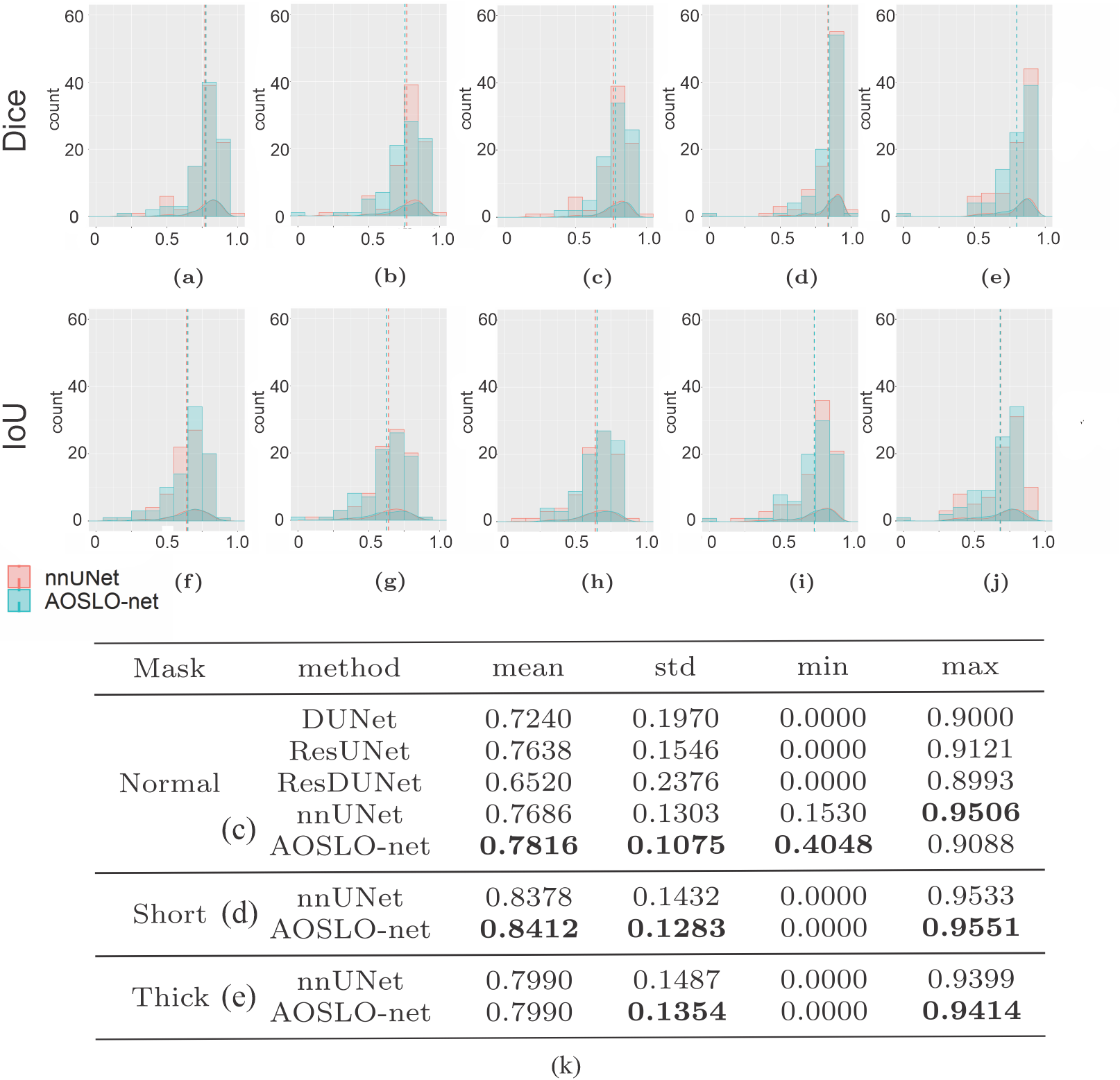}
    \caption{\textbf{Figure~\ref{Figure:Results_Statistics}. Performance metrics of AOSLO-net and nnUNet as well as other deep-learning models on segmenting the AOSLO images.} Histogram and density distribution of Dice score (a-e) and IoU (f-j) obtained from AOSLO-net and nnUNet when they are trained using (a,f) perfusion map and  normal mask set, (b,g) enhanced AOSLO images and  normal mask set, (c,h) two-channel MA images and normal mask set, (d,i) two-channel MA images and short mask set, (e,j) two-channel MA images and thick mask set.
    The result of AOSLO-net is denoted by cyan color whereas the result of nnUNet is denoted by red color. The \textbf{dash lines} denote the mean values of the Dice or IoU distributions.
    \textbf{(k)} The overall performance (mean value), performance stability (standard deviation), the worst and best case performance (minimal and maximum value) of Dice score of different AI models in segmenting AOSLO images.}\label{Figure:Results_Statistics}
\end{figure}

First, we train the AOSLO-net and nnUNet using perfusion map (Fig.~\ref{Figure:Data_Type_Input}(c)), enhanced AOSLO (Fig.~\ref{Figure:Data_Type_Input}(d)) and two-channel images (Fig.~\ref{Figure:Data_Type_Input}(e)), respectively, to examine which image modality can optimize the model performance. The normal mask set (Fig.~\ref{Figure:Data_Type_Input}(f)) is used as the target in these training processes. Our results in Figs.~\ref{Figure:Results_Statistics}(a-c,f-h) and Fig.3 in SI show that higher Dice and IoU scores are achieved for both nnUNet and AOSLO-net models when two-channel images are used, suggesting that two-channel images could provide more MA geometrical information to the segmentation models than the other two image modalities. Therefore, in the following section, we use the two-channel image set as the input of segmentation models to assess the model performance on different mask sets, including normal mask set, short mask set and thick mask set. The performance of AOSLO-net and nnUNet on these three mask sets is summarized in Fig.~\ref{Figure:Results_Statistics}, which shows that based on Dice and IoU, AOSLO-net achieves more high-quality predictions and fewer low-quality predictions for these three groups of masks.  Specifically, when models are trained with the normal and short mask sets, as shown in Figs.~\ref{Figure:Results_Statistics}(c,d), AOSLO-net achieves mean Dice scores of 0.7816 and 0.8412, respectively, which are higher than 0.7686 and 0.8378 of nnUNet. Additionally, the performance of AOSLO-net appears to be more stable than nnUNet, given smaller standard deviations in the results of AOSLO-net. When these two models are trained with thick mask set (see Fig.~\ref{Figure:Results_Statistics}(e)), although the mean Dice scores of AOSLO-net and nnUNet are the same, AOSLO-net preserves a stable performance with a relatively smaller standard deviation in Dice score. In addition to nnUNet, as shown in Fig.4 in SI, we specifically compare the performance of  AOSLO-net with other popular CNN-based models, i.e., Deformable UNet\cite{jin2019dunet}, ResUNet\cite{Zhang2017ResUNet,Yakubovskiy:2019} and Deformable ResUNet\cite{dai2017deformable,Zhang2017ResUNet,Yakubovskiy:2019}, using the normal mask dataset. The statistics of the Dice scores listed in Fig.~\ref{Figure:Results_Statistics}(k) suggest that AOSLO-net outperforms these models significantly in mean Dice score and standard deviation. 

\begin{figure}[H]
    \centering
    \includegraphics[width=1.0\linewidth]{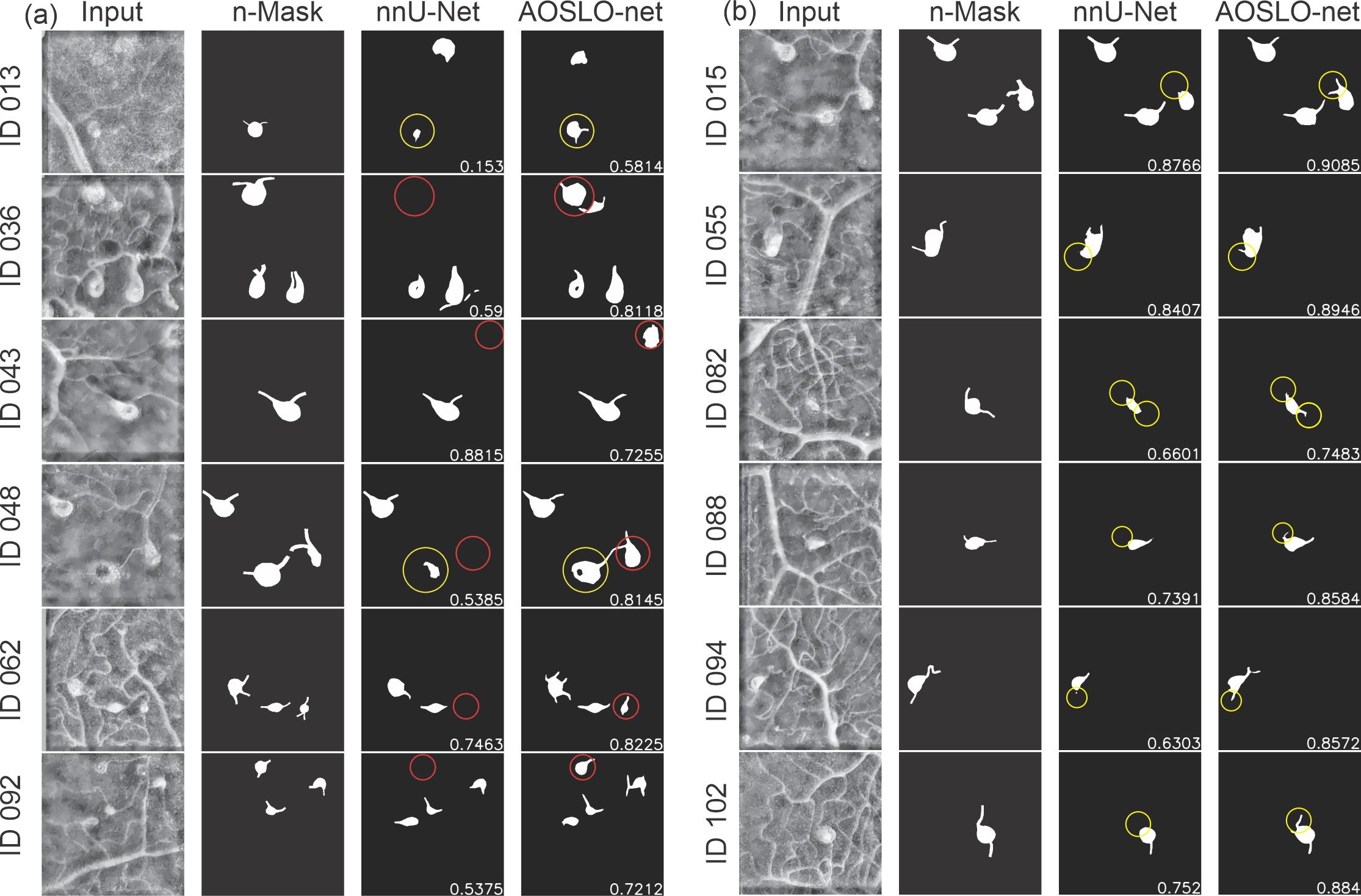}
    \caption{\textbf{Figure~\ref{Figure:Results_Normal_MoreMAs}. Typical examples of  AOSLO-net outperforming nnUNet on detecting (a) the MA body and (b) parenting vessels of MAs when they are trained using normal mask set (n-Mask)}. \textbf{First column}: enhanced AOSLO images. \textbf{Second column}: normal masks used to train the models. \textbf{Third column}: segmentation results of nnUNet. \textbf{Fourth column}: segmentation results of AOSLO-net. Numbers in images are the Dice scores.
    \textbf{(a)} For images with IDs 036, 043, 048, 062, and 092, the AOSLO-net is able to detect MAs, which are missed by nnUNet (marked in red circles). 
    For images with IDs 013 and 048, the AOSLO-net can better reconstruct the full shape of MAs compared to those of nnUNet (marked in yellow circles). We also note that in cases of IDs 013, 043 and 092, the segmentation models can even detect some potential MAs, which are not marked in the masks. \textbf{(b)}  For images with IDs 015, 055, 082, 088, 094 and 102, the AOSLO-net is capable of detecting both feeding and draining vessels connected to the MAs, while the nnUNet may miss one or both of the parenting vessels (marked in yellow circles).  These results show that the AOSLO-net is more reliable in detecting MA parenting vessels from the input images, which are essential in the MA classification --- a downstream task for disease diagnosis. }\label{Figure:Results_Normal_MoreMAs}
\end{figure} 

\subsection*{Morphology of MA body and its feeding and draining vessels}
To further examine the capabilities of AOSLO-net and nnUNet in extracting the detailed MA features, we perform image-wise analysis by comparing individual pairs of masks and model predictions. We focus on the model performance on detecting MA bodies and their feeding and draining vessels, respectively. Typical examples of the comparisons between nnUNet and AOSLO-net on detecting MA bodies are illustrated in Fig.~\ref{Figure:Results_Normal_MoreMAs}(a). We note that  AOSLO-net can identify the targeted MAs and extract MA bodies that are comparable with the masks, particularly for images containing multiple MAs, like IDs 036, 048, 062 and 092. In contrast, nnUNet fails to detect some of the MAs (highlighted in red circle).  As for the vessel detection, Fig.~\ref{Figure:Results_Normal_MoreMAs}(b) shows that nnUNet may ignore one or both of the parenting vessels of MAs (highlighted in yellow circle), while AOSLO-net can segment out these missed vessels (from complicated background with numerous vessels), which is essential for further morphological analysis.

\begin{figure}[H]
    \centering
    \includegraphics[width=0.98\linewidth]{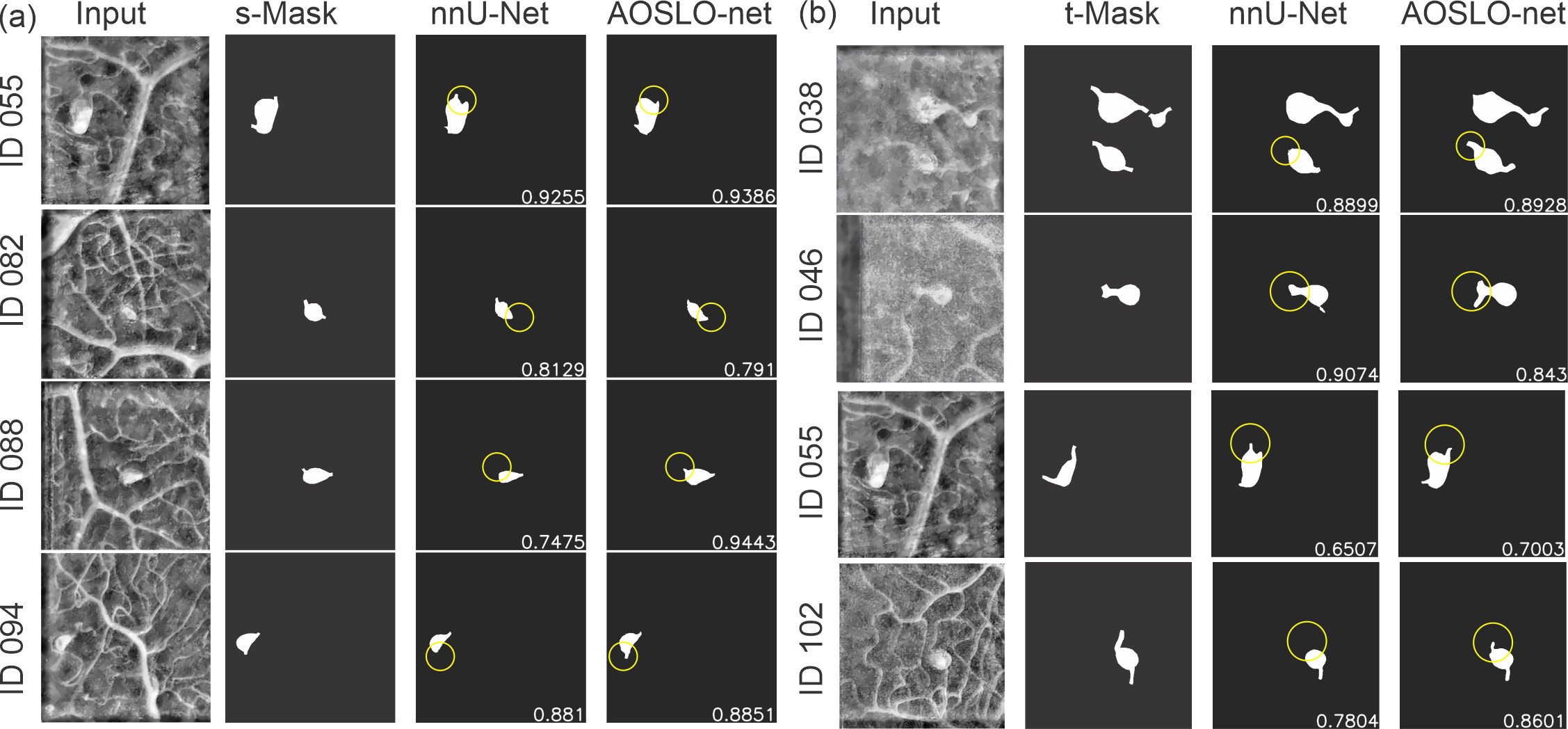}
    \caption{\textbf{Figure~\ref{Figure:Results_Short_Vessels_Diff}. Typical examples of  AOSLO-net outperforming nnUNet on detecting the parenting vessels of MAs when they are trained using (a) short mask set and (b) thick mask set.}  \textbf{First column}: enhanced AOSLO images. \textbf{Second column}: normal masks used to train the models. \textbf{Third column}: segmentation results of nnUNet. \textbf{Fourth column}: segmentation results of AOSLO-net. Numbers in images are the Dice scores. 
    (a) AOSLO-net are capable of detecting the parenting vessel of MAs in images with \textbf{ID 055, 082, 088 and 094}, whereas nnUNet misses one of the parenting vessels(marked in yellow circles). Particularly, nnUNet again mistakenly predicts the vessel on the upper left for the case of \textbf{ID 055}. (b) AOSLO-net are capable of detecting the parenting vessel of MAs in images with \textbf{ID 038, 046, 055 and 102}, whereas nnUNet misses one of the parenting vessels(marked in yellow circles). In case of \textbf{ID 055}, implementation of thick mask set improves the segmentation results of AOSLO-net, but not for  nnUNet.}\label{Figure:Results_Short_Vessels_Diff}
\end{figure}

 Next, we use another two sets of masks: one with shorter vessels compared to the normal dataset and one with thicker vessels, as learning targets to train AOSLO-net and nnUNet and compare their performance on detecting the feeding and draining vessels of MAs. We first train the model with short vessel mask dataset. The predictions from the two models in Fig.~\ref{Figure:Results_Short_Vessels_Diff}(a) show that the performances of nnUNet and AOSLO-net on detecting feeding and draining vessels are both compromised, as the shorter vessel masks provide less vessel end information. However, the AOSLO-net still can detect correct vessels in these four cases whereas nnUNet fails to predict some of the vessel ends (highlighted in yellow circle). These comparisons indicate that AOSLO-net is more robust against vessel length of the training masks. 
 
 When trained with thick vessel mask set, as shown in Fig.~\ref{Figure:Results_Short_Vessels_Diff}(b), AOSLO-net outperforms the nnUNet although the predictions of nnUNet have improved in detecting the feeding and draining vessels of MAs.  Fig.~\ref{Figure:Results_Short_Vessels_Diff}(b) shows that while nnUNet misses the MA bodies and vessel ends connected to MAs (highlighted in yellow circle), AOSLO-net can detect these feeding and draining vessels.  These results again demonstrate the robustness of AOSLO-net when trained with varying vessel thickness.
 
\subsection*{Morphology Quantification for the MA segmentation maps}
\begin{figure}[H]
    \centering
    \includegraphics[width=0.98\linewidth]{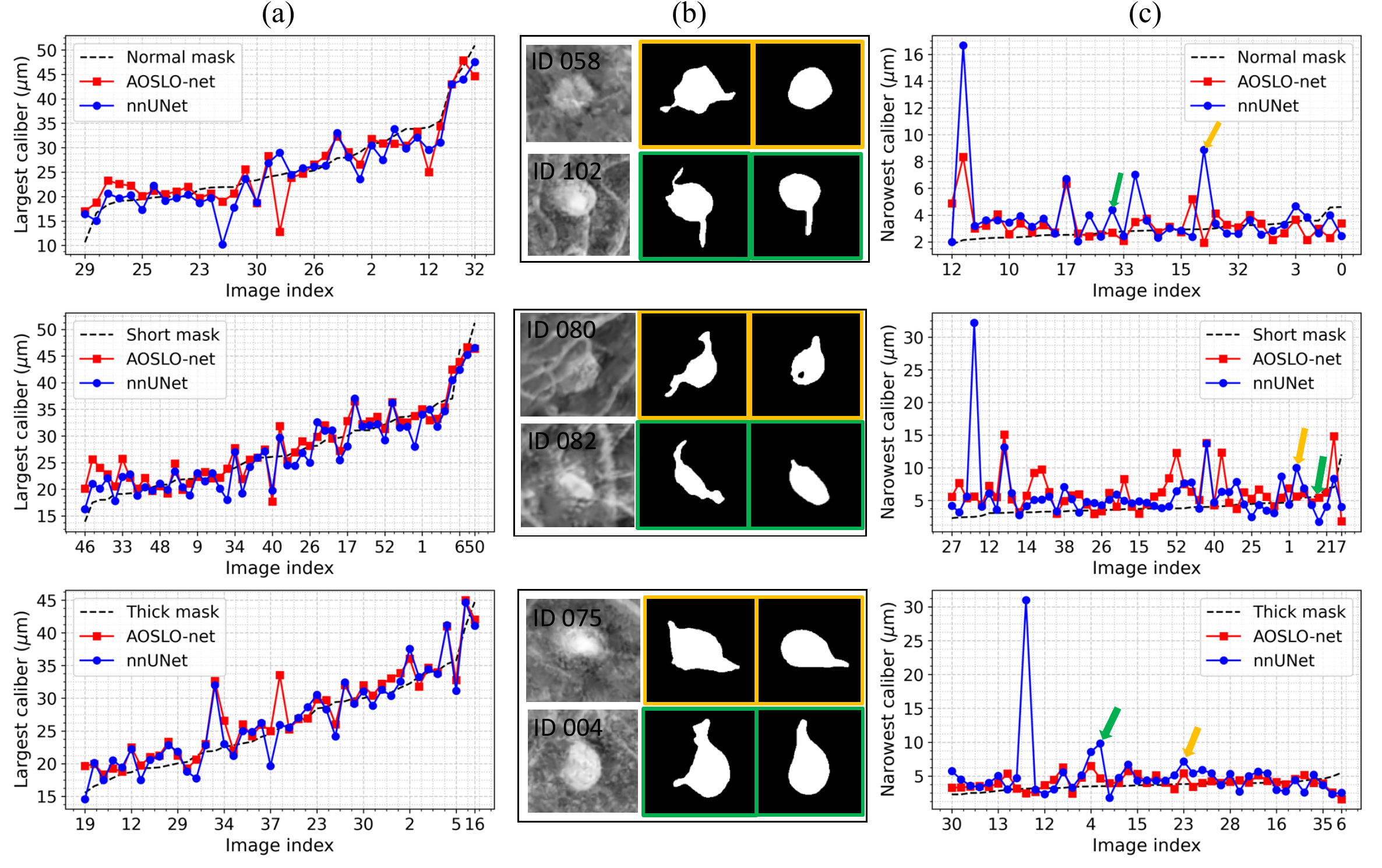}
    \caption{\textbf{Figure~\ref{Figure:Results_Morph_Params}. Quantification results for the largest caliber and narrowest caliber factors for the segmented MAs predicted by AOSLO-net and nnUNet trained with three sets of MA masks (normal, short and thick masks).} \textbf{(a)} The largest caliber (LC) results for the segmented MAs. The black dashed lines represent the reference MA LC values for different MAs, while the red and blue curves are the MA LC quantification results based on the segmentation maps from AOSLO-net and nnUNet using three different masks (from top to the bottom row). \textbf{(b)} Examples for the original enhanced perfusion maps (1st column), segmented MAs using AOSLO-net (2nd column) and nnUNet (3rd column) trained with different MA masks. The MA segmentation results in yellow and green bounding boxes correspond to the images in \textbf{(c)} marked with yellow and green arrows. \textbf{(c)} Narrowest caliber (NC) quantification results for segmented MAs using AOSLO-net and nnUNet trained with three different MA masks; the black dashed lines represent the reference MA NC values computed using the original 3 types of MA masks; the red and blue curves represent the NC quantification results using segmentation maps obtained by AOSLO-net and nnUNet using three different masks (from top to the bottom row).}\label{Figure:Results_Morph_Params}
\end{figure}

In order to better evaluate the MA segmentation performance of AOSLO-net and nnUNet, we compute three important MA morphological indices (``largest caliber'' - $LC$, ``narrowest caliber'' - $NC$ and the ``body-to-neck ratio'' - $BNR$) defined in~\cite{bernabeu2018estimation} to quantify the specific morphological characteristics for the segmented MAs. Specifically, we first compute the MA skeleton (or medial-axis) for every single MA using the ``Scikit-image'' package~\cite{van2014scikit}, and then apply the Euclidean distance transformation to compute the medial radius distances $D = \{d_i| i = 1,2,..., N\}$ ($N$ is the number of points on the MA skeleton) from all points of the MA skeleton to the background pixels (i.e., pixels of each MA contour). Consequently, the $LC$ value for each single MA corresponds to twice of the largest distance value in the sorted distance list $D_{sorted}$. Due to the varied vessel lengths of different MAs, we calculate the NC value for each single MA by selecting the 10 smallest medial radius distances from $D_{sorted}$, and double the average medial radius distances as the final $NC$ value. Based on the $LC$ and $NC$ values, the BNR value for each MA can be computed by using $LC/NC$. Fig.~\ref{Figure:Results_Morph_Params} gives some examples of the LC and NC quantification results for the segmented MAs predicted by AOSLO-net and nnUNet trained with three different MA masks (normal, short and thick mask sets). From Fig.~\ref{Figure:Results_Morph_Params}(a) and (c), we can find that the AOSLO-net trained with thick MA ground truth masks attains the best MA segmentation performance; the NC quantification results (red curve) as shown in the third row of Fig.~\ref{Figure:Results_Morph_Params}(c) are very close to the reference NC values (black dashed line) obtained from the thick masks. More details about the corresponding examples of the enhanced MA perfusion maps and the MA segmentation results using AOSLO-net and nnUNet are, respectively, shown in columns 1-3 of Fig.~\ref{Figure:Results_Morph_Params}(b). AOSLO-net can effectively detect the important small vessels for the heterogeneous MAs (see the 2nd column in Fig.~\ref{Figure:Results_Morph_Params}(b)), which play a very important role in different downstream tasks, e.g., MA morphological parameter quantification (NC, BNR, convexity), MA severity stratification, and in hemodynamics simulations. The three blue dots with very high NC values in different rows of Fig.~\ref{Figure:Results_Morph_Params}(c) are the same MAs for which nnUNet fails to detect the small vessels. Quantification results for the BNR of MAs based on the analysis of NC and LC is summarized in Fig.9 in SI.

\section*{Discussion and Summary}
Although 2D fundus photography has been primarily employed for DR screening and severity grading, other advanced imaging modalities, such as AOSLO, which is currently used for disease investigation and primarily for research purposes, can provide additional information regarding the retinal microvascular pathology, such as monitoring of variations in blood flow rates~\cite{couturier2015capillary}, detection and identification of the MA morphologies~\cite{dubow2014classification}, enhanced visualization of MA thrombus status~\cite{bernabeu2018estimation}, etc. The information from this imaging technique could be potentially used not only to improve the future accuracy of DR screening, but also to better predict the rate of DR worsening, with the goal of eventually providing individualized management and treatment intervention plans~\cite{goh2016retinal}. Due to the rapidly rising global prevalence of diabetes and shortage of skilled graders for retinal images, implementation of automatic screening techniques is desirable to accommodate the corresponding increasing need to screen and evaluate patients with diabetes ocular complications~\cite{he2020artificial,rajalakshmi2020impact}. Development of an automated segmentation technique for AOSLO images, which are at much smaller scale than fundus photographs, represents a unique challenge and provides the opportunity to elucidate the steps of \textit{in vivo} worsening and regression of MAs, a key lesion in DR.

In the last decade, DCNN has become the state-of-the-art technique in processing  retinal images. Although existing DCNN models have achieved high accuracy in analyzing fundus images, they have not been tested on retinal images from other modalities. In this work, we design AOSLO-Net to automatically segment MA bodies with their feeding and draining vessels from AOSLO images. We use a very deep model to extract high-level features to understand the MA morphology, which can vary substantially in shape and size between individual MAs, and to create segmentation based on that information. Our results show that the deep network structure can segment MAs from AOSLO images with high accuracy for both metrics and morphology. 

Our AOSLO dataset has a limited size, containing less than 100 videos, so the deep structure of AOSLO-net may cause overfitting, and it can be hard to train. Therefore, it is necessary to perform data augmentation according to specific dataset. Moreover, the MAs are of different shapes and many of them are not similar to others. This requires the AOSLO segmentation to be a composition of many few-shot problems. Therefore, data augmentation is necessary to train the deep network and fully utilize the data. To optimize the data augmentation, we use horizontal and vertical flip, rotation in uniformly distributed angles and scaling to achieve maximum space configuration of MAs. As a result, the AOSLO-net can learn MA geometric information very effectively from the limited dataset. In addition, transfer learning is desirable for training large and deep networks. In the current study, we use EfficientNet-b3~\cite{tan2020efficientnet} as the feature extractor (encoder), and pretrain AOSLO-net with ImageNet dataset~\cite{deng2009imagenet}, which is believed to accelerate the training convergence and is widely deployed in transfer learning based image analysis.

The relationship between model performance metrics and morphological features of MAs is also investigated carefully in our work. Accurate detection of feeding and draining vessels of MAs is critical to determine MA morphology, which is important to predict  blood flow characteristics and estimate the likelihood of thrombosis within the MA~\cite{dubow2014classification}. Since the areas of the end of the parenting vessels are much smaller compared to the areas of the whole MAs,  regional loss, like Dice and IoU, possibly ignore these vessel ends, and place too much emphasis on the MA bodies. However, correct segmentation of MA bodies alone is not sufficient for further classification. Thus, metrics like Dice and IoU are important but should not be the only metrics to evaluate the performance of MA segmentation. Our results demonstrate that AOSLO-net not only achieves high scores on these two metrics, but also captures the detailed features of the feeding and draining vessels of MAs, and therefore promises to be an effective MA segmentation method in the clinic.

\section*{Methods}
\subsection*{Image pre-processing}
\label{Section:Image_Preprocessing}
As shown in Fig.~\ref{Figure:Data_Type_Input}(b), the raw AOSLO images are featured with intense background noise and low contrast. Thus, we perform imaging pre-processing on these raw images and generate multi-modality images to improve the effectiveness of the training. First, we generate a set of perfusion maps by tracing the blood flow in the MAs and micro-vessels using the pixel-by-pixel standard deviation method on different frames of the AOSLO video~\cite{bernabeu2018estimation}. As shown in Fig.~\ref{Figure:Data_Type_Input}(c), the vessels with blood motion appear bright in the perfusion map, while static tissue shows up as dark background. To improve the quality of the perfusion maps,  we apply the following methods to denoise and enhance the images: (1) we employ \textbf{fast non-local means method} to remove the background noise; (2) use \textbf{normalization} to make the pixel value lie in the interval \([0, 1]\); (3) apply contrast limits adaptive histogram equalization (\textbf{CLAHE}) to enhance image contrast without over stretching the contrast in specific areas and balance the overall contrast; (4) apply \textbf{Gamma Correction} to remove some bright stripes on the background surrounding tissues caused by CLAHE. The impact of each of these four pre-processing steps on the raw perfusion maps is illustrated in Fig.~\ref{Figure:Appendix_Method_Preprocessing} in SI.

 Perfusion maps may not be able to accurately illustrate the geometries for all MAs as thrombosis may occur in some of the MAs, which leads to presence of nonperfused areas~\cite{bernabeu2018estimation,li2020predictive}. Therefore, we have developed enhanced AOSLO images to provide more details on the boundaries of each MA, as shown in Fig.~\ref{Figure:Data_Type_Input}(d). The procedure of creating enhanced AOSLO images follows three main steps: (i) taking the average over all the frames in MA video; (ii) reversing image colour; (iii) performing local mean filtering. We note that in some enhanced AOSLO images, the boundaries of MAs and their parenting vessels are not clearly illustrated due to the low quality of the AOSLO images.  Therefore, we further generate a two-channel image set by concatenating perfusion map and enhanced AOSLO images (see Fig.~\ref{Figure:Data_Type_Input}(e)), which use the information of the blood flow inside MAs to compensate the missing information of the MA boundaries.

Due to the limited size of the AOSLO data set, we employ data augmentation to increase the number of images for training the AOSLO-net. We apply three types of transformations, including flip, rotate and scaling, to the AOSLO images (perfusion map, enhance AOSLO and two-channel images) and their corresponding masks.  The augmentation procedure follows three steps: (1) images are \textbf{flipped horizontally and vertically} with probability of 0.5; (2) flipped images are \textbf{rotated} with angles in the set
\begin{equation*}
    \{0, \frac{2\pi}{N}, \frac{4\pi}{N}, \ldots, \frac{2(N-1)\pi}{N}\}
\end{equation*}
where \(N\) is selected to be 32. (3) The rotated images are \textbf{scaled} with a factor randomly selected between 0.7 and 1.4 to improve the robustness of AOSLO-net on segmenting MAs with varying sizes. Typical examples of augmented images are shown in Fig.~\ref{Figure:Appendix_Method_Augmentation} in SI.

\subsection*{Architecture of AOSLO-net and network training}

Inspired by the popular UNet structure~\cite{ronneberger2015u}, AOSLO-net is composed of two key parts: encoder and decoder. The function of the encoder is to extract the features of MAs at different levels whereas the decoder integrates these extracted features to compose the segmentation results. Since the role of the encoder is critical to the performance of the segmentation model, we adopt the current SOTA image classification network EfficientNet-b3~\cite{tan2020efficientnet} with a depth of 5 as the encoder in the AOSLO-net.  We also apply transfer learning in AOSLO-net through pre-training the EfficientNet-b3 using ImageNet~\cite{deng2009imagenet} to achieve quick convergence during training. The evolution of the loss in the training process can be found in  Figs.~\ref{Figure:Appendix_Model_Loss_Record} and~\ref{Figure:Appendix_Model_Scheduler} in SI. 

The pre-processed AOSLO images are split into 5 folds. While one fold is reserved as test data, the rest 4 folds are used to train and validate the AOSLO-net. We perform a 10-fold cross-validation using these 4 folds of images, meaning that these images are further separated into 10 folds, with 9 folds used for training after augmentation and one fold used for validation. Overall, 1600 augmented images are used to train the AOSLO-net. 10 images are used for validation and 17 images are used to test the model performance.

The loss function is a combination of binary cross-entropy and Dice loss,
\begin{equation*}
    Loss = BCE + \alpha \times Dice,
\end{equation*}
where BCE and Dice are defined as
\begin{equation*}
    \begin{aligned}
        BCE  & = \sum_n\left[ y_n \cdot \ln x_n + (1 - y_n) \cdot \ln (1 - x_n) \right] \\
        Dice & = \frac{2|X\cap Y|}{|X|+|Y|}.
    \end{aligned}
\end{equation*}
Here, \(|X|,\ |Y|\) are model prediction and target respectively, \(x_n,\ y_n\) are pixel values  in \(|X|\) and \(|Y|\); \(\alpha \) is set to 0.2 to equalize the contribution of BCE and Dice to the loss. We also introduce Hausdorff distance, which is commonly used to describe the contour difference of two shape, into the loss function, but it does not improve the segmentation results (see Fig.6 in SI).

We use the Adam optimizer with learning rate of 0.001 and weight decay of~$10^{-8}$. During training,  a Plateau scheduler is specified such that once the validation loss does not decrease for \(5\) epochs,  the learning rate is decreased to \(1/10\) of the current value to facilitate the convergence of the AOSLO-net. 
The training is initially set to last 200 epochs, but it may end earlier due to the implementation of Plateau scheduler. The batch size is set to 16.

\subsection*{Post-Processing and Ensembling}

\noindent
\textbf{Binarization.} The pixel value of the outputted images from AOSLO-net lies within range [0, 1] resulting from employing the sigmoid activation function in the last layer of AOSLO-net. To quantify the geometries of MAs, we convert these pixel values into a binary form, meaning that the pixel value is either 1 (belongs to an MA), or 0 (not MA). A threshold of 0.5 is applied to binarize the segmented images in the current study. 
\\
\textbf{Clearing.} We note that some segmented images contain small fragments that are mistakenly predicted as MAs. Thus, we specify an area threshold of 1024 pixels, below which the fragments are removed from the segmented images. \\
\textbf{Ensembling.} Following the work of~\cite{isensee2021nnu}, we employ ensembling method by selecting three best models, out of the 10 trained models, based on their performance on the validation set and perform a union of their outputs to improve the model performance. Some examples illustrating the effect of post-processing are shown in Fig.~\ref{Figure:Appendix_PostProc_BinClean} in SI.

\subsection*{Performance metrics}
We evaluate the segmentation performance of AOSLO-net and other segmentation models using the Dice coefficient and intersection of union (IoU), defined as follows,
\begin{equation}
    \begin{aligned}
        Dice & = \frac{2|X\cap Y|}{|X|+|Y|},  \\
        IoU  & = \frac{|X\cap Y|}{|X\cup Y|},
    \end{aligned}
\end{equation}
where \(|X|\) and \(|Y|\) denote the matrix representations of the target image and the corresponding prediction image, respectively.

\subsection*{Data availability}
The dataset used in the current study is not currently available publicly, but it will be released upon request.

\subsection*{Code availability}
The code is not currently available publicly but could be released upon  publication of this manuscript.

\paragraph{Duality of Interest.} No potential conflicts of interest relevant to this article were reported.

\paragraph*{Acknowledgements.} Q.Z., M.X., S.C., Y.D., H.L. and G.E.K. acknowledge the support from R01 HL154150 and U01 HL142518. K.S. and J.K.S. acknowledge the support by NEI 5R01EY024702-04 as well as grants from Research to Prevent Blindness, JDRF 3-SRA-2014-264-M-R, and the Massachusetts Lions Eye Research Fund.  High Performance Computing resources were provided by the Center for Computation and Visualization at Brown University and the Extreme Science and Engineering Discovery Environment (XSEDE), which is supported by National Science Foundation grant numbers ACI-1053575, TG-DMS140007 and TG-MCB190045.

\paragraph*{Author Contribution.} Q.Z., K.S., M.X., S.C., Y.D., H.L., J.K.S. and G.E.K. contributed to the study concept and design. Q.Z., K.S., S.C., Y.D. and H.L. contributed to acquisition, analysis, or interpretation of data. Q.Z., K.S., M.X., S.C., Y.D., H.L., J.K.S. and G.E.K. contributed to drafting of the manuscript. Q.Z., M.X., S.C., Y.D., H.L. and G.E.K. contributed to algorithm development. Q.Z. and M.X. performed hardware implementation. Q.Z. and M.X. contributed to statistical analysis. J.K.S. and G.E.K supervised the project.


\appendix
\section{Data}
\subsection{Preprocessing on Perfusion Map}
\begin{figure}[H]
    \centering
    \includegraphics[width=\linewidth]{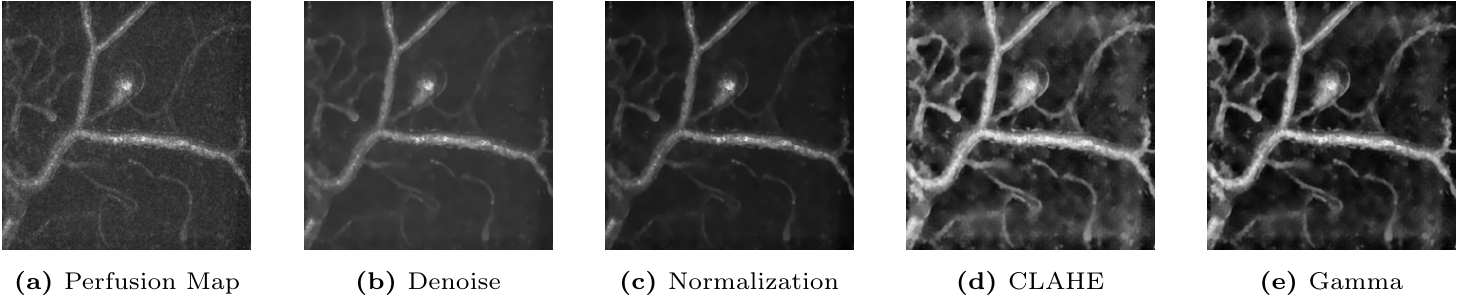}
    \caption{\textbf{Figure~\ref{Figure:Appendix_Method_Preprocessing}. An example of the sequential image preprocessing steps for perfusion map.} \textbf{(a)} Raw perfusion map, which contains background noise and suffers from low image contrast. \textbf{(b)} Perfusion map after applying the non-local means filtering method, which creates a denoised image. \textbf{(c)} Perfusion map after intensity normalization in the range of \([0, 1]\), which enhances the contrast. \textbf{(d)} Perfusion map after employing contrast limited adaptive histogram equalization (CLAHE). \textbf{(e)} Perfusion map after employing Gamma correction. We applied these sequential preprocessing steps for all the perfusion maps 
    in AOSLO-net training and testing.  }\label{Figure:Appendix_Method_Preprocessing}
\end{figure}

\subsection{Data augmentation}
\begin{figure}[H]
    \centering
    \includegraphics[width=\linewidth]{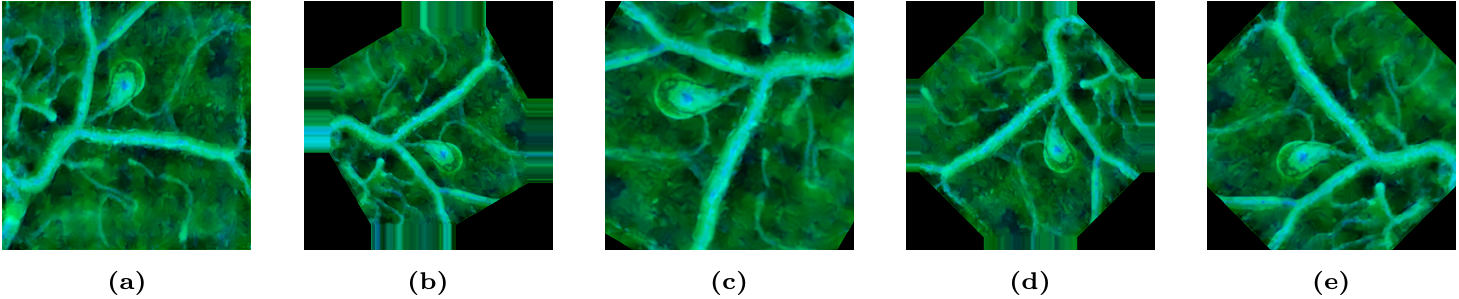}
    \caption{\textbf{Figure~\ref{Figure:Appendix_Method_Augmentation}. Examples of data augmentation.} \textbf{(a)} Two-modality image (using the processed perfusion map and the enhanced AOSLO image as two channels in RGB space). \textbf{(b)-(e)} Augmented images from (a) by performing different transformations: 
    \textbf{(b)} flipping vertically, rotating with an angle $\theta={\pi}/{6}$ and scaling with a 
    factor $\lambda<1$; \textbf{(c)} flipping horizontally, $\theta={\pi}/{3}$ and $\lambda>1$; \textbf{(d)} flipping both horizentally and vertically, $\theta={\pi}/{4}$ and $\lambda<1$; \textbf{(e)} no flipping, $\theta={3\pi}/{4}$ and $\lambda\sim 1$. The rotating angle was uniformly distributed, while the scaling factor was randomly selected in an appropriate range when we performed data augmentation on our AOSLO dataset.
    }\label{Figure:Appendix_Method_Augmentation}
\end{figure}

\begin{figure}[H]
    \centering
    \includegraphics[width=\linewidth]{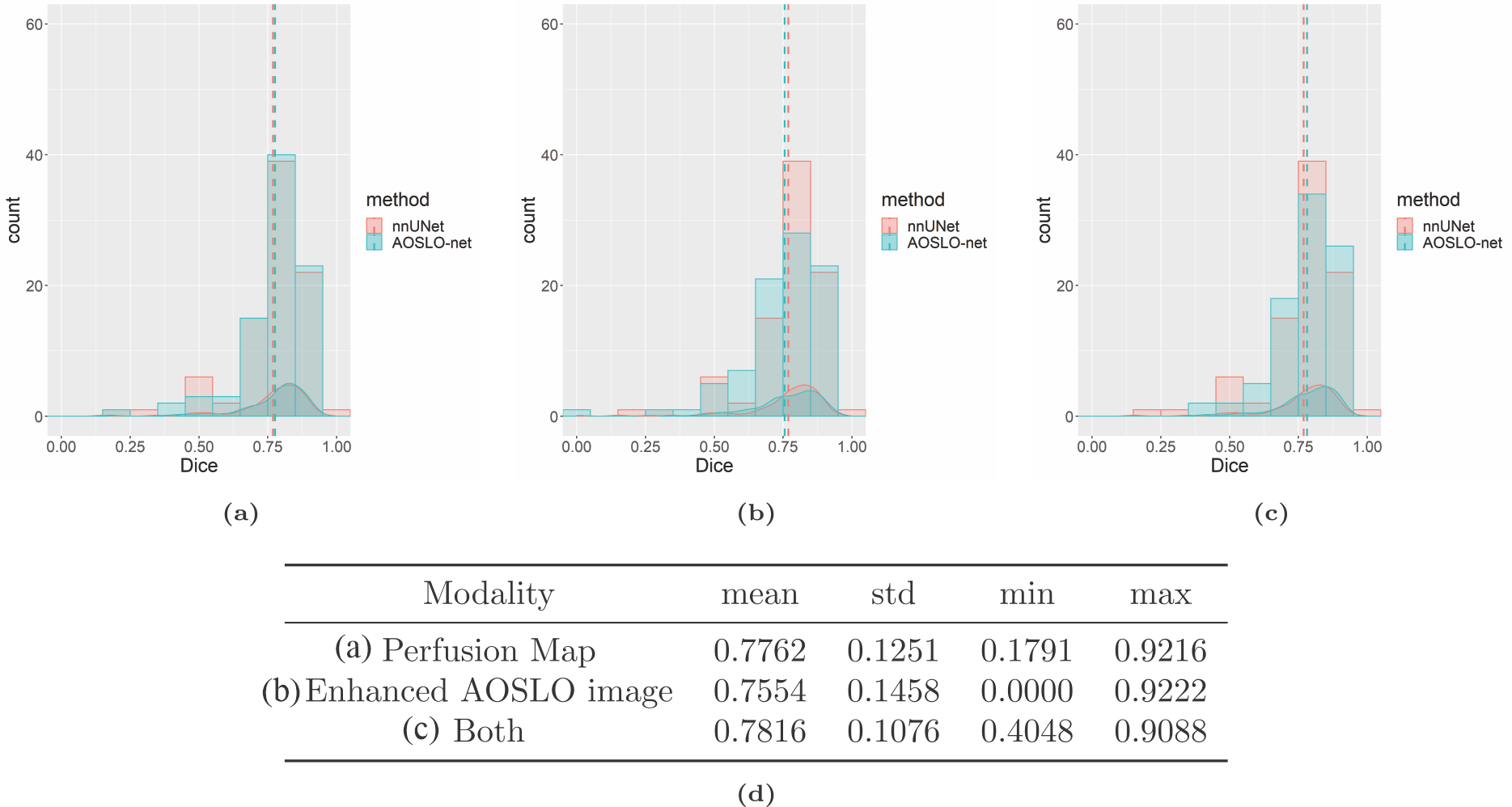}
    \caption{\textbf{Figure~\ref{Figure:Appendix_Method_Modality}. Performance of AOSLO-net and nnU-Net with different input modalities. } \textbf{(a)} Histogram and density distribution of Dice score obtained by using the perfusion maps as input. The result of AOSLO-net is denoted by cyan color whereas the result of nnU-Net is denoted as red color. The mean Dice score (denoted by dash line) of AOSLO-net is slightly higher than that of nnU-Net. 
    \textbf{(b)} Histogram and density distribution of Dice score obtained by using the enhanced images as input. In this case, the mean Dice score of AOSLO-net is slightly lower than that of nnU-Net. Moreover, the AOSLO-net is more likely to make low-quality predictions. \textbf{(c)} Histogram and density distribution of Dice score obtained by using two-modality images as input, where the AOSLO-net performs better than the nnU-Net. 
    \textbf{(d)} The statistics of AOSLO-net based on different input modalities. Using two-modality images as input results in the best performance. 
    }\label{Figure:Appendix_Method_Modality}
\end{figure}

\section{AOSLO-net model}
\subsection{Network}
\begin{figure}[H]
  \centering
  \includegraphics[width=\linewidth]{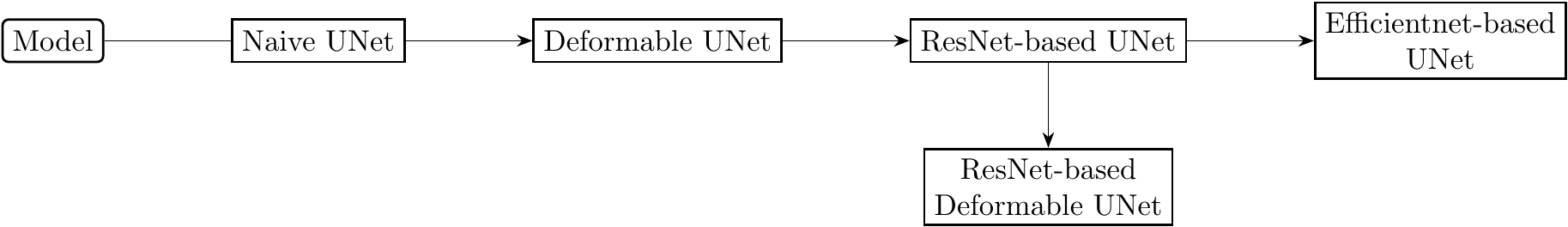}
  \caption{\textbf{Figure~\ref{Figure:Appendix_Model_Network_History}. A list of segmentation models that have been tested in the current work.} The arrow points to the model with a better performance in terms of Dice and IoU score.
  }\label{Figure:Appendix_Model_Network_History}
\end{figure}
\noindent\textbf{UNet.} The vanilla version of UNet\cite{ronneberger2015u} was proposed in 2015. This network tends to erroneously predict the background vessels as MAs, and overall the predicted MA shapes are not satisfactory.
\\
\textbf{Deformable UNet.} Deformable UNet\cite{jin2019dunet} uses the deformable convolution layer instead of the original convolution layer. Deformable UNet shows improved performance on distinguishing the MAs from the vessels, but it fails to capture the MA shape and vessels connected to MAs.
\\
\textbf{ResUNet.}  We implemented ResUNet with \verb|Segmentation Models Pytorch|\cite{Yakubovskiy:2019}, a python package. ResNet101 was used as backbone, which included more parameters than the official implementation. As a result, this UNet is able to better capture the MA shape and the inlet outlet vessels. However, the overall performance of this UNet is not sufficient for MA shape quantification.
\\
\textbf{ResNet-based Deformable UNet.} The main body of this kind of UNet is the same as the ResNet-based UNet, but the convolution layers are replaced to be deformable ones. We implemented it by substituting ordinary convolution kernel in \verb|Segmentation Models Pytorch| with a deformable convolution kernel\cite{dai2017deformable}. But these modifications do not improve model performance on segmenting AOSLO images. 
\\
\textbf{EfficientNet-based UNet.}  Using EfficientNet-b3\cite{tan2020efficientnet}, a SOTA classification network, as encoder in AOSLO-net  has show significant improvements over the models listed above. We use the implementation from \verb|Segmentation Models Pytorch|.

\subsection{Loss}
Binary CrossEntropy (BCE) is the most common loss function for classification and segmentation. However, in our study, there is also a strong imbalance since the MA areas are relatively small compared with the whole image. Under these circumstances, the BCE loss was overwhelmed and the neural network predicted MAs as background. To overcome this imbalance, we use regional loss based on Dice and IoU to maximize the intersection ratio of the prediction and the ground truth. 

\begin{figure}[t]
  \centering
  \includegraphics[width=0.6\linewidth]{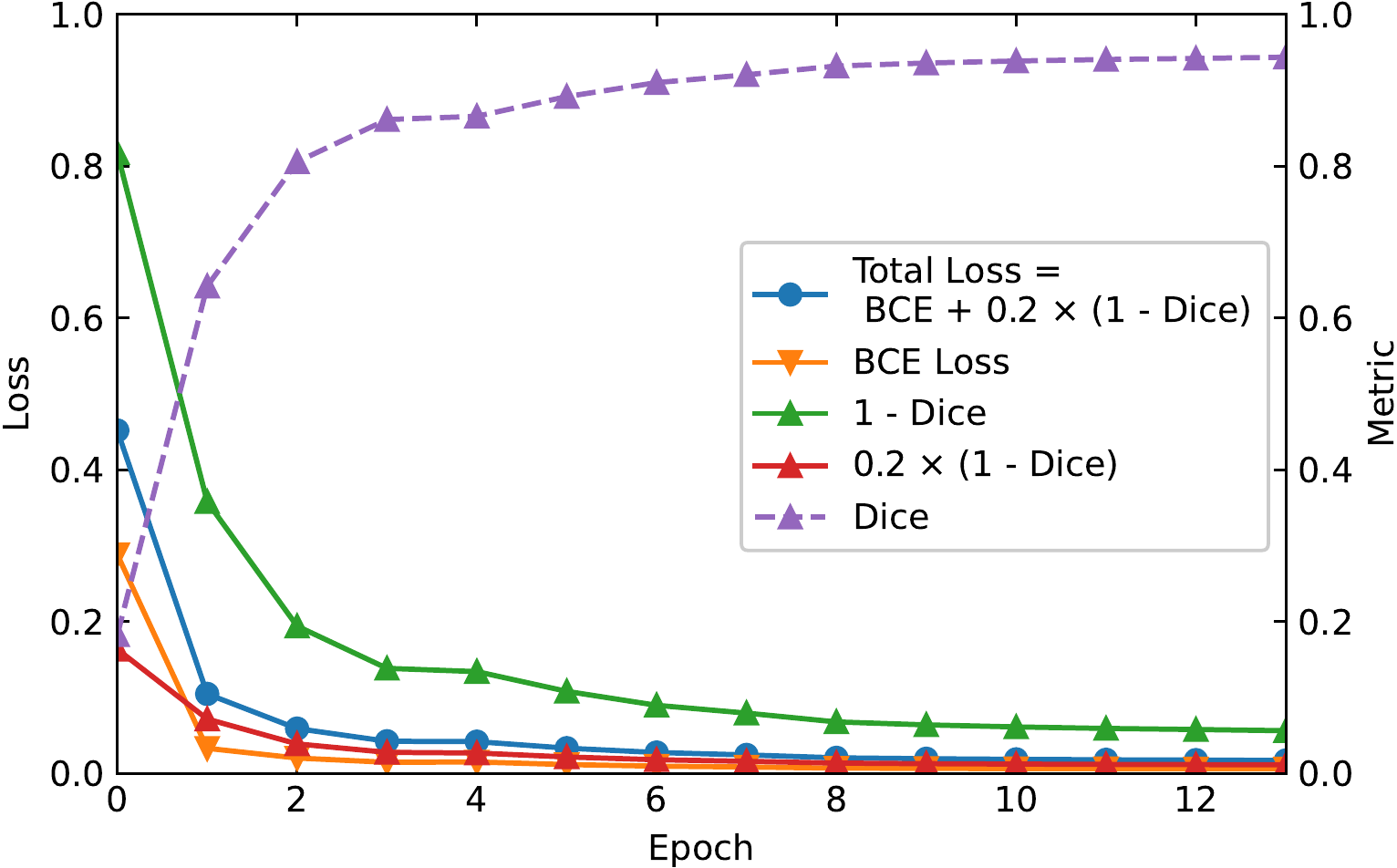}
  \caption{\textbf{Figure~\ref{Figure:Appendix_Model_Loss_Record}. Training Losses and Dice score of AOSLO-net versus training epochs.} The original Dice loss (in green color), which is defined as $1 - \text{Dice}$, is much larger compared to the binary cross entropy (BCE) loss (in orange). In order to balance the contribution of two terms in the total loss function, a weighting coefficient $0.2$ was applied for the 
  Dice loss. The convergence of the rescaled Dice loss, $0.2\times(1 - \text{Dice})$, is shown in red color, while the total loss is shown in blue.
 }\label{Figure:Appendix_Model_Loss_Record}
\end{figure}

Despite the advantage of regional loss like Dice and IoU, they may cause oscillation of the training loss curve. Thus, we added BCE as a smoothing loss to the regional loss. We used a coefficient \(\alpha=0.2\) to equalize contribution of BCE and regional loss as shown in Figure~\ref{Figure:Appendix_Model_Loss_Record}.

\begin{figure}[H]
    \centering
    \includegraphics[width=0.8\linewidth]{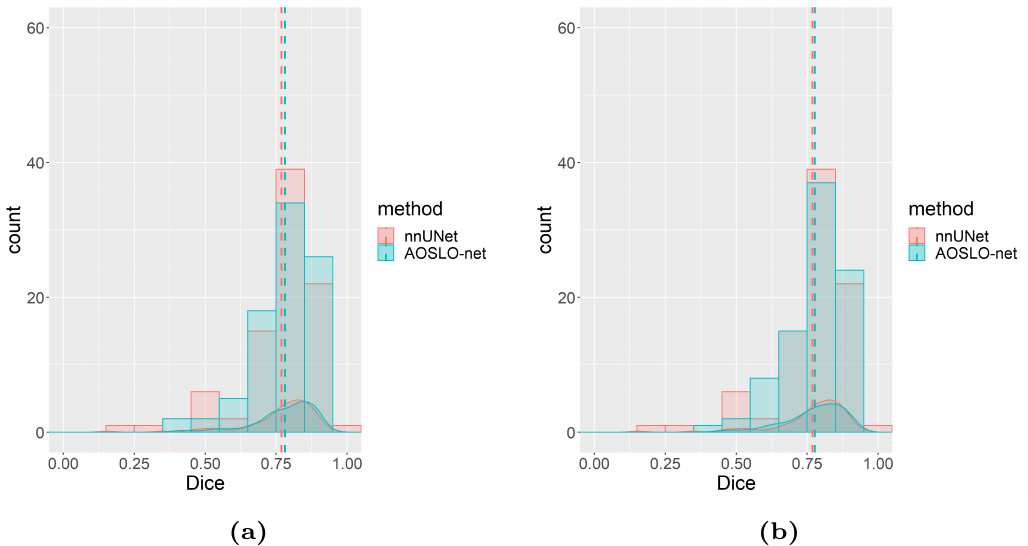}
    \caption{\textbf{Figure~\ref{Figure:Appendix_Model_Loss_Hausdorff}. Performance of AOSLO-net and nnU-Net with and without Hausdorff distance loss.} \textbf{(a)} Histogram and density distribution of Dice score when the models are trained with Hausdorff distance as part of the loss.  \textbf{(b)} Histogram and density distribution of Dice score when the models are trained without using the Hausdorff distance. 
    We observe that applying the Hausdorff distance did not improve the performace of AOSLO-net. Instead, the mean value of Dice score dropped from 0.7816 to 0.7774 for the AOSLO-net.}\label{Figure:Appendix_Model_Loss_Hausdorff}
\end{figure}

We also tested the Hausdorff distance, which is defined as
\begin{equation}
    d_{H}(X, Y) = \max \{\sup_{x\in X}d(x, Y), \sup_{y\in Y}d(y, X)\},
\end{equation}
where \(X\) is prediction and \(Y\) is target, \(d\) is metric of the space, which is the Eucleadian distance in our case. The Hausdorff distance is used to describe the contour difference of two shape. However, as shown in Fig.~\ref{Figure:Appendix_Model_Loss_Hausdorff}, the effect of using the Hausdorff distance as part of the loss was not notable to our current method.

\subsection{Optimizer}
\begin{figure}[H]
  \centering
  \includegraphics[width=0.6\linewidth]{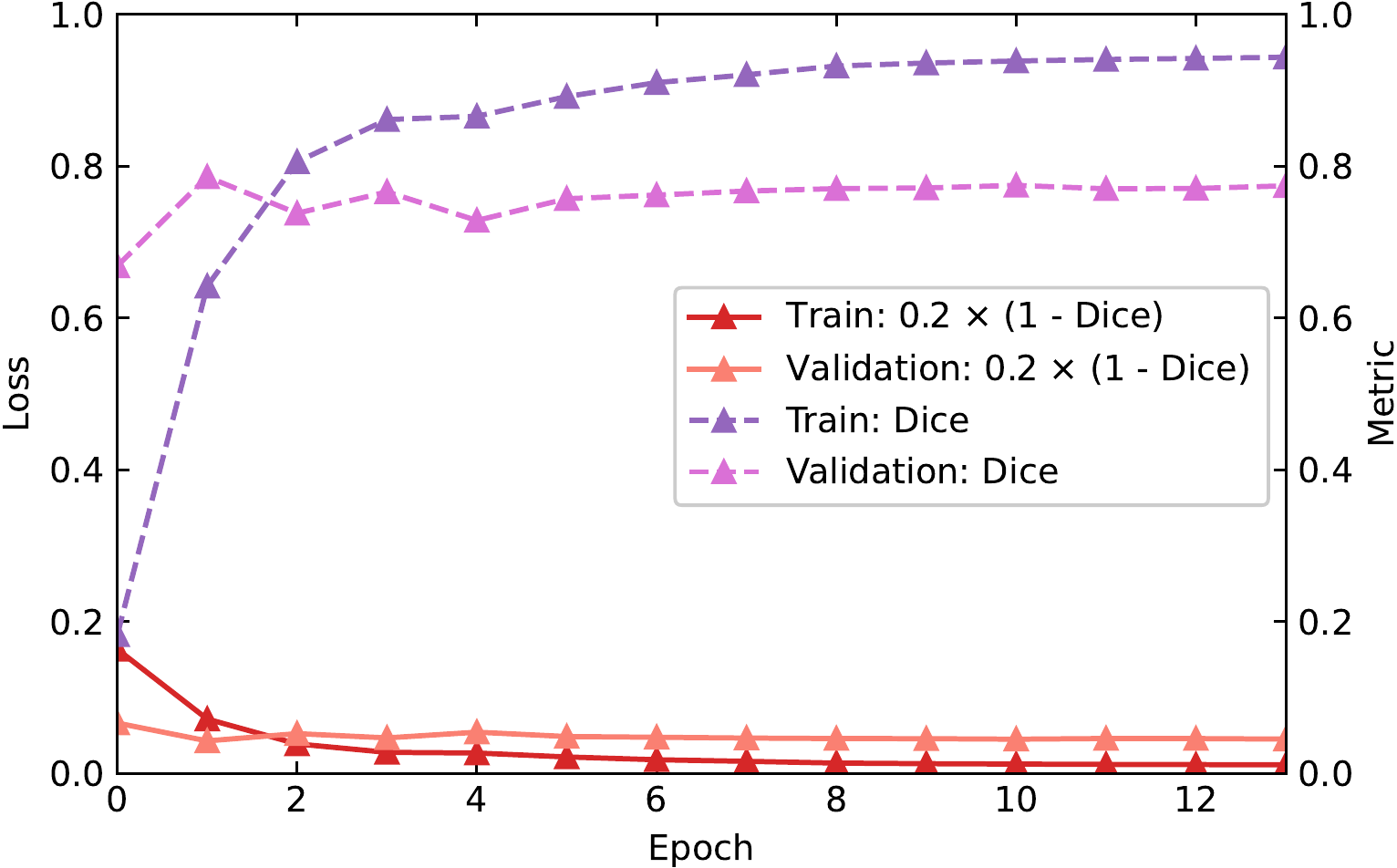}
  \caption{\textbf{Figure~\ref{Figure:Appendix_Model_Scheduler}. Loss and Dice score of AOSLO-net in training and validation.} The AOSLO-net undergoes a quick convergence and achieves a good validation Dice score upon convergence.
  }\label{Figure:Appendix_Model_Scheduler}
\end{figure}

We first used the SGD as optimizer, the default for many deep learning based computer vision problems and we found that loss descending rate is very slow. Thus, we employ Adam optimizer and the training usually converged in 20 epochs. More importantly, after using Adam as optimizer, the segmentation performance on validation and test set is also improved.
\\
In order to prevent the overfitting, we used the Plateau learning rate scheduler. When the validation loss does not decrease for \(5\) epochs, the scheduler decreases the learning rate to \(1/10\) of the current one, so the network converges before overfitting. The effect of using the scheduler is illustrated in Figure~\ref{Figure:Appendix_Model_Scheduler}, where the training finished in less than 15 epochs, while nnU-Net will use 1000 epochs by default.

\section{Post Processing}
\subsection{Binarization and Clearing}

\begin{figure}[H]
    \centering
    \includegraphics[width=0.8\linewidth]{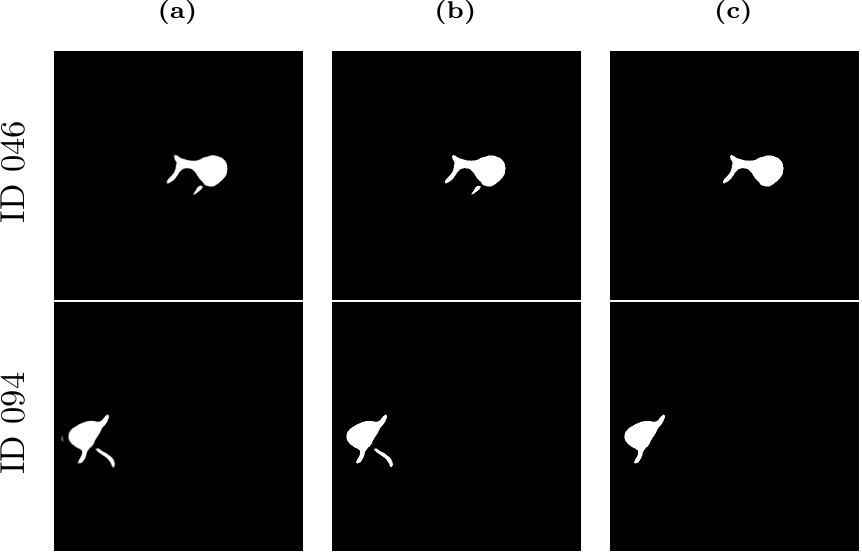}
    \caption{\textbf{Figure~\ref{Figure:Appendix_PostProc_BinClean}. Two examples of using image binarization and clearing to remove vessels that are not connected to MAs.} \textbf{(a)} The raw output image of AOSLO-net. \textbf{(b)} Output image after binarization. \textbf{(c)} Output images after clearing. }\label{Figure:Appendix_PostProc_BinClean}
\end{figure}

\subsection{Morphology quantification results for the BNR factor of MAs}
\begin{figure}[H]
    \centering
    \includegraphics[width=0.98\linewidth]{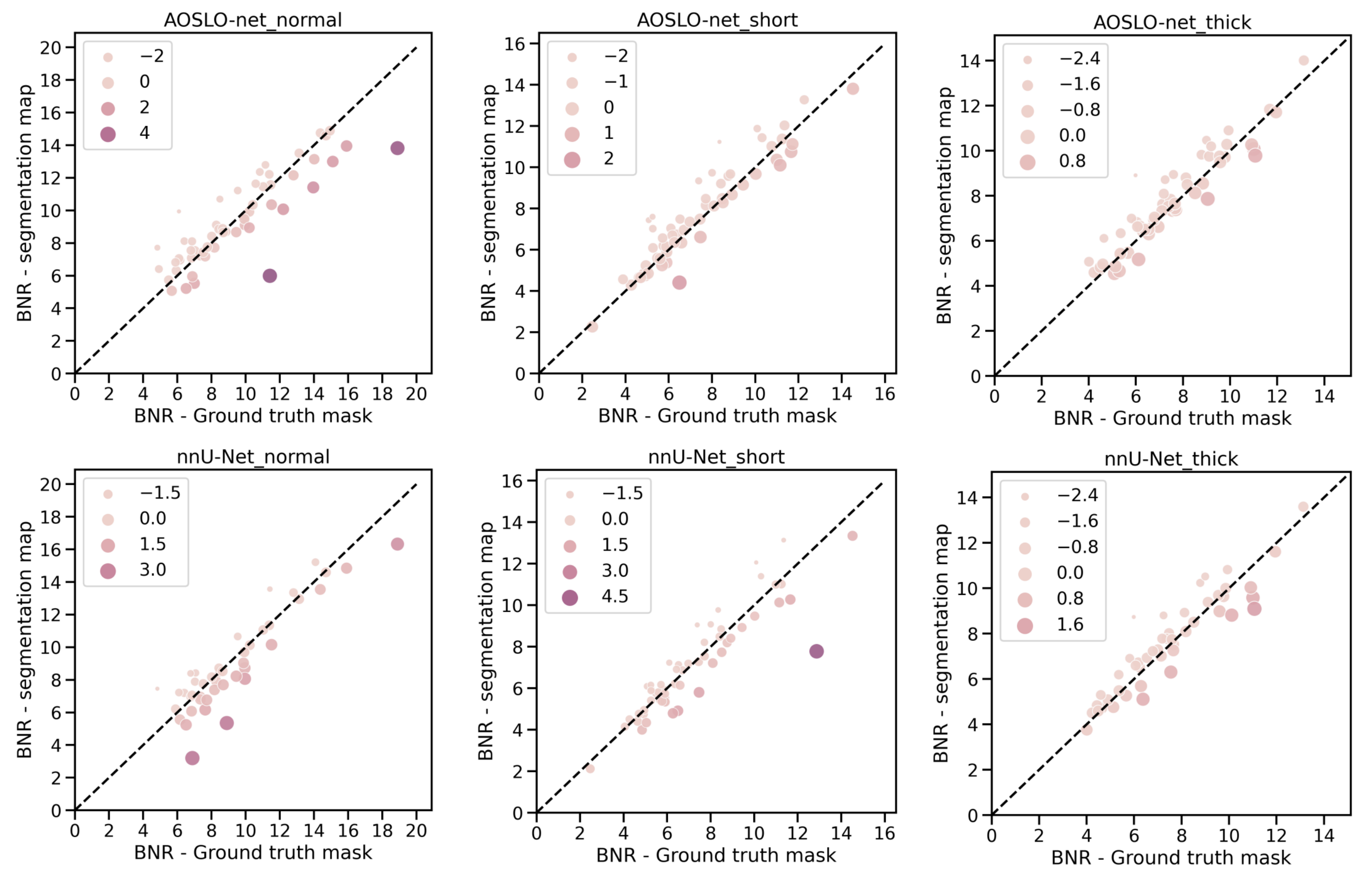}
    \caption{\textbf{Figure~\ref{Figure:Results_BNR}. Quantification results of the BNR (body-to-neck ratio) factor for the segmented MAs using AOSLO-net and nnU-Net trained with normal, short and thick masks.} \textbf{First row}: Comparisons of BNR quantification results based on AOSLO-net segmentation maps and the corresponding adopted three different types of MA masks (normal, short and thick masks as shown in columns 1--3); \textbf{Second row:} Comparisons of BNR quantification results based on nnU-Net segmentation maps and the corresponding adopted MA masks (normal, short and thick masks as shown in columns 1--3). The dot size and color represent the estimated BNR value difference between using the MA segmentation maps and the original ground truth masks. AOSLO-net segmentation maps enable to generate more accurate BNR estimation results (BNR difference values are relatively smaller) in comparison to nnU-Net segmentation maps. Moreover, AOSLO-net trained with thick masks can capture more accurate the MA geometry features (especially for the narrowest calibers). The corresponding BNR estimation results are closer to the BNR values estimated by using the normal masks.}\label{Figure:Results_BNR}
\end{figure} 
\end{document}